Thermo-magnetic properties of the magnetocaloric layered materials based upon FeMnAsP: a Green function-method approach

Osvaldo F. Schilling

Departamento de Física, Universidade Federal de Santa Catarina, Campus, Trindade, 88040-900, Florianopolis, SC. Brazil.

Abstract

The compounds $FeMnAs_xP_{1-x}$ are very promising as far as commercial applications of the magnetocaloric effect are concerned. However, the theoretical literature on magnetocaloric materials still adopts simple molecular-field models in the description of important properties like the entropy variation that accompanies applied isothermal magnetic field cycling, for instance. We apply a Green function theoretical treatment for such analysis. The advantages of such approach are well-known since the details of the crystal structure are incorporated in the model, as well as a precise description of correlations between spins of the transition metal ions can be obtained. For the sake of simplcity we adopt a simple one-exchange parameter Heisenberg model, and the observed first-order phase transitions are reproduced by the introduction of a biquadratic term in the hamiltonian. Good agreement with experimental magnetocaloric data for $FeMnAs_xP_{1-x}$ compounds is obtained, as well as an agreement with the magnetic field dependence for these properties predicted from the Landau theory of continuous phase transitions.

1. Introduction

Since the discovery of the giant magnetocaloric effect by Pecharsky and Gschneidner [1] several families of materials have been investigated for commercial use as a replacement for conventional refrigerating systems based upon gas-liquid transitions under pressure cycling. The most promising of such materials display ( almost) abrupt FM or AFM transitions close to room temperature with narrow associated hysteresis loops[2]. Such first order ( or close to first order, i.e., sharp) transitions are associated with the best cooling capabilities, measured in terms of the adiabatic temperature variation associated with magnetic field cycling, and also the isothermal entropy variation with magnetic field cycling. The direct measurement of such temperature and entropy variations is experimentally difficult in comparison to the measurement of the magnetic moment, both as a function of temperature or external applied magnetic field[3]. Therefore, reversible thermodynamic relations involving the measured magnetization have been adopted to calculate entropy variations and evaluate performance for new materials. The project of new structures on the other hand requires theoretical models to correlate experimental data with for instance exchange parameters associated with different ion arrangements in the actual materials. Theoretical studies have been carried out in which the main objective has been the calculation of thermo-magnetic properties from a given model of the considered material. Such treatments employ effective( molecular)-field theory(MFT)[3,4], an approach well known for its relative mathematical simplicity as well as for its limitations, mostly related to oversimplification. Among the widely exploited virtues is the possibility of obtaining reasonable results for thermodynamic and magnetic properties which in many cases would be impossible to get otherwise. The defects include the impossibility of including the details of the crystal structure in the model, the prediction of excessively high values for FM or AFM phase transition temperatures, and the prediction of null magnetic correlations and specific heat above the transition in zero applied magnetic field, just to mention the most relevant for the present purposes[5]. Therefore, it would be important if a more precise theoretical technique could be applied, preferably with a level of mathematical complexity similar to that of MFT, to calculate  the magnetic properties close to the transition.  In particular, any theoretical model should be able to yield the temperature and magnetic field dependence of the magnetic entropy. From the Gibbs-Helmholtz expression for the Free Energy of an array of magnetic moments it is possible to obtain an expression for the system entropy which is a function of the correlation function between the moments, which appears in the expression of the Heisenberg hamiltonian. A rather simple and adequate approach for statistical-mechanical calculations of correlations is provided by the double-time Green functions technique[6]. Such technique is suitable to deal with many-body systems since it is focused on the calculation of correlation functions and their propagation in time and space, which is exactly what is needed in interacting magnetic spin lattices. The method can be turned "static" by taking the same values for the initial and final instants of time in the propagation, and the Fourier transform on the position parameters of the lattice leads directly to the values of correlation functions between spins as required in theoretical magnetism. A wealth of literature on the Green function applications to the Heisenberg model exists, and we cite only those more relevant for the present study[6-12]. In the next section 2 we develop the main expressions to be used in the simulations. In particular, we adopt the random phase approximation (RPA)[6] in order to break up at an early stage a sequence of increasingly complex combinations of spin operators. We are going to specialize on layered structures of the HCP symmetry, which is the case of the $FeMnAs_xP_{1-x}$ structures, in which Fe and Mn are stacked in hexagonal layers along the c axis of a structure with As and P in between the transition metal atoms in each layer. Figure 1 shows the structure. The Green function method permits the detailed consideration of the lattice structure since structure factors get into the calculations, which does not happen in MFT. It has been determined [13]that the magnetic transition in this family of compounds involves a redistribution of electrons around Fe ions which substantially affects the lattice parameters, although the cell volume remains practically unaltered. Following a procedure adopted by Lines and Jones[9], we consider such an effect by inserting a dependence of the exchange parameter upon the square of the magnetization. This is similar from the practical point of view to the explicit consideration of a biquadratic term in the hamiltonian, which has been inserted in the past in MFT to reproduce such elastic effects upon the transition[3,4]. The weight of such a term is phenomenologicaly taken as the minimum necessary to produce a non-hysteretic abrupt transition at zero applied magnetic field. In view of the previous work of Tegus[3] and others on this family of compounds, we assume all ions have spin *S*=1 ( and *g* factor of 2) in agreement with the average number of about 4 Bohr magnetons experimentally observed per formula unit. In section 3 we compare the numerically calculated values of isothermal entropy changes under magnetic field cycling with experimental values available in the literature. We also calculate the adiabatic temperature variations under magnetic field cycling. Recent theoretical work on the application of the Landau theory of phase transitions to the magnetic transitions in magnetocalorics has focused on the dependence of the isothermal entropy variation upon the magnetic field strength[14]. Although such theory should apply to continuous transitions we obtain a similar dependence even for the discontinuous transitions considered.

2. Green functions theory for the correlations between spins in FeMnAsP structures.

The Fe and Mn sublattices stacked along the c-axis in Figure 1 are considered similar in magnetic terms in order to restrict the model to one single fitting parameter, the exchange constant *J*, which will be taken as the same for all nearest-neighbor ions( 6 in the same plane and 6 in the neighbor planes along the c axis). The parameter *J*(*s*) includes a correction proportional to the mean magnetization *s* ( the same for both Fe- or Mn- sublattices) squared in the form *J*(*s*)= $J(1 - \alpha s^2)$. The "weight" $\alpha$ in the present case is chosen phenomenologically as 0.4 since it is approximately the least value for which a discontinuous magnetic transition is obtained, with no hysteresis( to reproduce experimental observations of very narrow hysteresis loops). It must be pointed out that for any value of the spin *S* different from ½ a dependence of the exchange energy upon higher powers of *s* is indeed expected on theoretical grounds even in the absence of magneto-elastic effects, and 0.4 is of the expected magnitude for a correction term quadratic in *s*. This will give rise to a "biquadratic" correction, as proposed in previous molecular-field theoretical treatments of this same structure.

The Heisenberg Hamiltonian for *N* ions is given by the equation

$$H = -Ng\mu_B Bs - 2J(s) \Sigma_{f,m} \boldsymbol{S}_f \cdot \boldsymbol{S}_m \qquad (1)$$

The first term gives the interaction of each spin with an applied magnetic field $B$, $g=2$, and $\mu_B$ is the Bohr magneton. We drop the index $s$ in the expression for $J$ for brevity. The second term accounts for the exchange energy due to nearest neighbor (f,m) interactions. According to Green functions theory ( which main details can be found in the references[6-12]) we represent static propagation across the lattice from points $g$ to the neighbor points $h$, $j$, by the function $\ll S_g^+, S_{j,h}^- \gg$ which obeys the equation

$$E \ll S_g^+, S_h^- \gg = \left(\frac{F}{2\pi}\right)\delta_{gh} - 2J(\Sigma_{xy}^j + \Sigma_z^j) \ll (S_g^z, S_j^+ - S_g^+, S_j^z), S_h^- \gg \quad (2)$$

In (2) $F$ stands for the thermal average taken over a canonical ensemble(represented by singly-pointed brackets below)of the commutator of the operators $S^+$ and $S^-$, as

$$F = <[S_g^+, S_g^-]> \quad (3)$$

the summations over x,y are carried out over the 6 neighbors in the same ( x,y) plane as the ion $g$, and summations over the z-direction on the 6 neighbors in the planes above and below. The last term on the right side of (2) produces an increasingly complex series of combinations of spin operators. Following the references [6-12]we replace $S_{j,g}^z$ in the correlation functions simply by the average value $s$ ( the random-phase aprroximation-RPA) which breaks the series of spin operator products in the first term.

The average value of the spin correlation is the main result to be obtained, and this will require that $\ll S_g^+, S_{j,h}^- \gg$ be Fourier transformed in the following way( in the notation of Lines[7])

$$\ll S_g^+, S_h^- \gg = \left(\frac{2}{N}\right)\sum_{\mathbf{K}} G_{1\mathbf{K}} e^{i\mathbf{K}\cdot(\mathbf{g}-\mathbf{h})} \quad (4a)$$

or

$$\ll S_g^+, S_h^- \gg = \left(\frac{2}{N}\right)\sum_{\mathbf{K}} G_{2\mathbf{K}} e^{i\mathbf{K}\cdot(\mathbf{g}-\mathbf{h})} \quad (4b)$$

provided either ( subscript 1) $g$ and $h$ are in the same plane( same sublattice, with $N/2$ atoms per sublattice in the crystal), or ( subscript 2) they are in different planes(different sublattices). The summations run over $N/2$ values of **K** in the first Brillouin-zone of each sublattice.

We replace (4a) and (4b) in (2), and perform the inverse Fourier transforms to obtain a couple of equations for $G_{1K}$ and $G_{2K}$:

$$G_{1\mathbf{K}}(E - \mu s + \lambda_{xy} s) = \left(\frac{F}{2\pi}\right) - \lambda_z s\, G_{2\mathbf{K}} \qquad (5a)$$

$$G_{2\mathbf{K}}(E - \mu s + \lambda_z s) = -\lambda_{xy} s\, G_{1\mathbf{K}} \qquad (5b)$$

where $\mu = 12J$, $\lambda_{xy} = J\sum_{xy}^{j=1\ldots 6} e^{i\mathbf{K}\cdot(\mathbf{j}-\mathbf{g})}$, $\lambda_z = J\sum_{z}^{j=1\ldots 6} e^{i\mathbf{K}\cdot(\mathbf{j}-\mathbf{g})}$. The energy $E$ already includes the Zeeman contribution $g\mu_B B s$. The $\lambda$ are the structure factors which characterize the specific lattice structure of the material, which is absent in MFT treatments. One immediately obtains

$$G_{1K} = \left(\frac{F}{2\pi}\right)\left(\frac{\lambda_2}{E-\mu s} + \frac{\lambda_1}{E-(\mu-\lambda_{xy}-\lambda_z)s}\right) \qquad (6a)$$

$$G_{2K} = \left(\frac{F}{2\pi}\right)\lambda_1\left(-\frac{1}{E-\mu s} + \frac{1}{E-(\mu-\lambda_{xy}-\lambda_z)s}\right) \qquad (6b)$$

where $\lambda_1 = \frac{\lambda_{xy}}{\lambda_{xy}+\lambda_z} = 1 - \lambda_2$. By means of equations (4) to (6) the correlation functions of two operators $A$ and $B$ are obtained[6-12]:

$$<BA> = \lim_{\epsilon\to 0} i\int_{-\infty}^{\infty} \frac{\ll A,B\gg(E=\omega+i\epsilon) - \ll A,B\gg(E=\omega-i\epsilon)}{e^{\omega/T}-1}\, d\omega \qquad (7)$$

in which we make use also of

$$\lim_{\epsilon\to 0}\left[\frac{1}{\omega+i\epsilon-E} - \frac{1}{\omega-i\epsilon-E}\right] = -2\pi i\delta(\omega - E) \qquad (8)$$

The numerical solution of this set of equations leads to the values of the exchange energy $E_{exch} = NJz(<S^+_i S^-_j> + <S^z_i S^z_j>)$ in which $z$ is the number of nearest neighbors and **i** and **j** are vectors designating neighbor sites[11]. From the limiting behavior expected for $<S^z_i S^z_j>$ at zero temperature and at the transition temperature $T_c$, we take $<S^z_i S^z_j> = s^2 + s(1-s^2)\phi_\delta$, since $<S^+_i S^-_j> = 2s\phi_\delta$ (from (3)), with $\boldsymbol{\delta} = \mathbf{j} - \mathbf{i}$ and

$$\phi_\delta = \sum_{k,n} \frac{\lambda_n \exp(i\mathbf{k}\cdot\boldsymbol{\delta})}{\exp\left(\frac{E_{k,n}}{T}\right)-1} \qquad (9)$$

In (9) there is a summation over $n=1,2$, and the two roots of the denominators in (6a) and (6b) are $E_{k,n}$ and give the dispersion relations for interaction energies[10].

In particular, the value of the mean value of the magnetization $s$ must comply with the self-consistent solution of the equation

$$\phi_{\delta=0} = \frac{2-3s+\sqrt{4-3s^2}}{6s} \tag{10}$$

which is valid for $S=1$[11].

The variation of the entropy $S$ under isothermal cycling of applied external magnetic fields $B$ at $T_c$ is a quantity of major interest in this work. From the calculation of the correlation functions and corresponding exchange energies between $B=0$ and $B=B$, for a sequence of field steps $dB$ it is possible to obtain numerically( for $T=T_c$)

$$\Delta S = \int_0^B \frac{dE_{exch}}{T} \tag{11}$$

And compare (11) with experimental results and theoretical predictions for temperatures close to first-order ( or almost first-order) transitions. We also calculate the isofield entropy variation beteen $T=0$ and a given $T$, shown below in Figures 2 and 3. It is also possible to obtain the adiabatic temperature variation under magnetic field cycling from the same calculations. It must be stressed that the method of Green functions is not in practice more difficult of implementing that those based upon molecular-field theory. Wolfram Mathematica codes have been written for such purpose.

3. Results

Figure 2 displays a typical set of calculations of isofield magnetic entropy variations( per ion) against the final temperature, for applied magnetic fields between 0 and 10 tesla, with temperature in units of the exchange constant $J$. As explained earlier we adopt $\alpha=0.4$ so that an almost discontinuous transition is obtained( see the sharp raise close to $T_c$ in the first plot on the left, for $B=0$). For zero applied field the entropy in absolute units is 0.71 at $T_c$. The result 12.1 for $T_c/J$ is consistent with the expected for a HCP structure( as compared with the related FCC lattice result which is 11.9[12]), and sets the value of $J$ at about 25K for a critical temperature of 300K. Figure 3 displays an analogous entropy variation calculation for the case $\alpha=0$, i.e., for a second-order transition, for $B=0$ and $B=1$ tesla. The entropy converges towards $\ln(3)=1.099$, the theoretical limit, at high temperatures. The entropy at $T_c$ in absolute units is 0.81. Notice that series expansions calculations for a FCC structure ( there is no available data for HCP) predicts the difference between the entropy at $T_c$ and that for complete randomness 1.099 for $S=1$ as 0.29 at zero magnetic field, which agrees with the Green function calculations shown in Figure 3( see arrow in the figure). The sharp raise displayed in Figure 2 also makes evident the advantage of cycling the field close to $T_c$ since the entropy variation with field is maximum for a sharp transition. Figure 3 shows that continuous transitions on the other hand produce entropy curves which almost superimpose each other for zero field and also for higher fields so that a much smaller magnetocaloric effect is observed.

Recent theoretical work [14] has focused on the application of Landau's theory of second order phase transtions to calculations of entropy variation close to $T_c$. A theoretical expression relating field variation $\Delta B$ with the obtained $\Delta S$ at $T_c$ has been proposed in the form $\Delta S = -a\,\Delta B^{2/3} + b\,\Delta B^{4/3}$ for continuous transitions[14]. Figure 4 shows that this formula applies to our Green function calculations as well, carried out for an abrupt transition at $T_c$. Here we adopt eq. (11) with fixed $T$ at $T_c$ and vary the field in small steps and integrate until reaching each final field indicated in the horizontal axis. The scale in this plot is consistent with that of Figure 2, and the effect of obtaining complete order with application of an infinite magnetic field is to make the total absolute entropy variation at $T_c$ converge to about 0.75, as expected. In Figure 5 we show a comparison between predicted values of $\Delta T$, the adiabatic temperature variation at $T_c$ obtained by cycling the magnetic field from 0 to a final field strength. There are data available( from Tegus PhD Thesis[3]) only for small field variations. Our calculations show that such temperature variations ( whose precise measurement is rather complicated to obtain) might be twice as large as those reported. It has been pointed out in an older study that $\Delta T$ should also be proportional to $B^{2/3}$ [15], and a dependence in the form $-a\,\Delta B^{2/3} + b\,\Delta B^{4/3}$ [14] is shown by the solid line fit to the theoretical points. In Figure 6 we compare our calculations for the entropy variation with Tegus's Thesis [3]results, and good agreement is obtained.

## 4. Conclusions

Layered structures containing Mn and Fe ions stacked in alternating hexagonal planes have been proven very promising as far as technical and commercial applications of the magnetocaloric effect are concerned. The theoretical treatment of the magneto-thermal properties of such structures has been attempted so far by means of molecular-field models, which recognizedly cannot take account of the geometrical details of any particular structure, as well as the correlations between neighboring ions which are averaged throughout. This paper is actually the continuation of previous work by the author [16] in which efforts to overcome the weaknesses of simple molecular-field treatments of such structures have been undertaken. In the previous work we applied the so-called constant coupling method. Such method considers in detail a pair of neighboring spin interactions and takes the average over the remaining interactions. We have shown that a method can indeed be developed in which nearest-neighbor interactions, as well as next-nearest-neighbors can be considered in the calculations, but the problem of a large number of scarcely known exchange interactions to be inserted in the equations remained. In the present paper we took the inverse approach of adopting the least possible number of fitting parameters, namely just one, $J$, but to improve the calculations potential accuracy by adopting double-time Green functions. When one adopts Green functions one focus upon the most relevant quantities for this particular problem, which are the correlation functions between spins, as stressed in section 2 of this paper. We developed a detailed Model based upon $S=1$ ( different) ions stacked in a HCP environment. The procedure is explained in detail and its implementation in the form of ( for instance) Mathematica codes is not essentially more difficult than that necessary for the constant-coupling method adopted earlier. Abrupt ( or almost first-order) transitions can be reproduced by inserting a "biquadratic" term phenomenologically ( there is no way of obtaining it precisely just by theoretical methods since the existence of such a term has both mathematical grounds-the spinor algebra related to the Pauli exclusion principle in exchange terms for spin 1 systems-and also it is related with magneto-elastic coupling, with the relative magnitudes of both unknown). We have actually had no difficulty of obtaining reasonable to good agreement with experimental results of the magnetocaloric effect quantities, with a single fitting parameter, *J*= 25K. We have been able also to show that the present method is in agreement with the field dependence of $\Delta S$ proposed for second-order transition materials like Gd.

Further work is underway to extend the method to other structures of interest for magnetocaloric applications.

References

[1] V.K.Pecharsky and K.A.Gschneidner, Phys.Rev.Lett. **78**, 4494 (1997).

[2] E.Brück, et al., Int.J.Refrig. **31**, 763 (2008).

[3] O.Thegus, PhD Thesis, University of Amsterdam (2003). Available at: www.science.uva.nl/research/mmm/Tegus/Tegusthesis.pdf( accessed on 14/12/2015).

[4] R. Zach, M. Guillot, and J. Tobola, J. Appl. Phys. **83**, 7237 (1998).

[5] J.S.Smart, Effective Field Theories of Magnetism, Saunders, NY, 1966.

[6] S.V.Tyablikov, Methods in the Quantum Theory of Magnetism, Plenum Press, NY, 1967.

[7] M.E.Lines, Phys.Rev. **133**, A841(1964).

[8] M.E.Lines, Phys.Rev. **131**, 540(1963).

[9] M.E.Lines and E.D.Jones, Phys. Rev. **139**, A1313(1965).

[10] A. Narath, Phys. Rev. **140**, A854 (1965).

[11] J.A.Copeland and H.A. Gersch, Phys. Rev. **143**, 236(1966).

[12] R.A. Tahir-Kheli, Phys. Rev. **132**, 689 (1962).

[13] N.H.Dung, et al., Adv. En. Mat. **1**, 1215 (2011).

[14] J. Lyubina, et al., Phys. Rev. **B83**, 012403 (2011).

[15] H. Oesterreicher and F.T. Parker, J. Appl. Phys. **55**, 4334 (1984).

[16] O.F. Schilling, Braz. J. Phys. **43**, 214 (2013).

Figure captions

Figure 1: Lattice structure of Fe-Mn-As-P family of compounds showing the arrangement of transition metals ions in neighbor planes. Red symbols: Fe ions. Blue symbols: Mn ions( from [16]).

Figure 2: Calculated absolute entropy ( maximum at 1.099 for $S=1$) variation $\Delta S$ per ion between $T=0$ and a final temperature, for $B=0$ and 2,4,6,8, and 10 tesla, for $\alpha=0.4$. A sharp transition is obtained at $T/J= 12.1$ and zero field.

Figure 3: Same simulations as in Figure 2 but for $\alpha=0$, continuous transition, for $B=0$ and 1 tesla. The curves almost coincide with each other. The arrow indicates the value 0.81 for the entropy variation at $T_c$ and zero field, which has the expected value as compared with series expansions calculations for the related FCC structure.

Figure 4: A complete series of simulated entropy variation values at $T_c$ up to very high fields-cycling, for $\alpha=0.4$. We obtain good agreement with the expressions proposed in [14] for continuous transitions.

Figure 5: Adiabatic temperature variation at $T_c$ for $\alpha=0.4$ compared with results for $MnFeP_{0.45}As_{0.55}$ on p.66 of Tegus 's Thesis [3].

Figure 6: Isothermal entropy variations upon magnetic field cycling at $T_c$ for $\alpha=0.4$ compared with results for $MnFeP_{0.45}As_{0.55}$ on p.64 of Tegus's Thesis [3]( entropy units J $kg^{-1}$ $K^{-1}$ as in [3]). The traced line is a guide to the eye.

Figure1

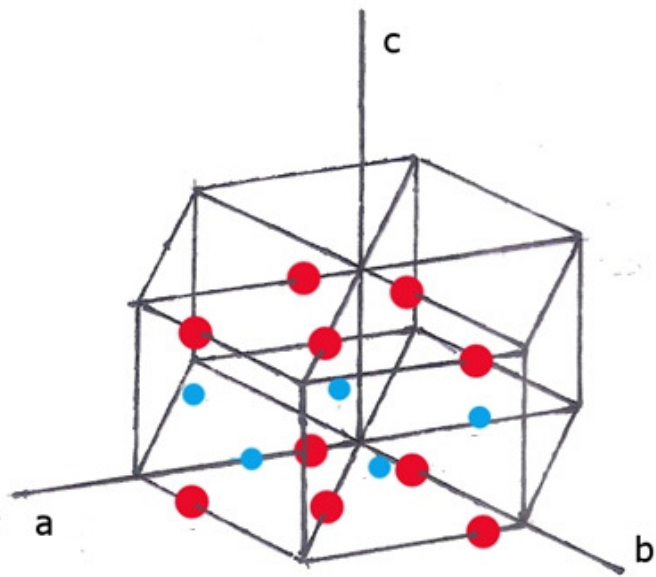

Figure 2

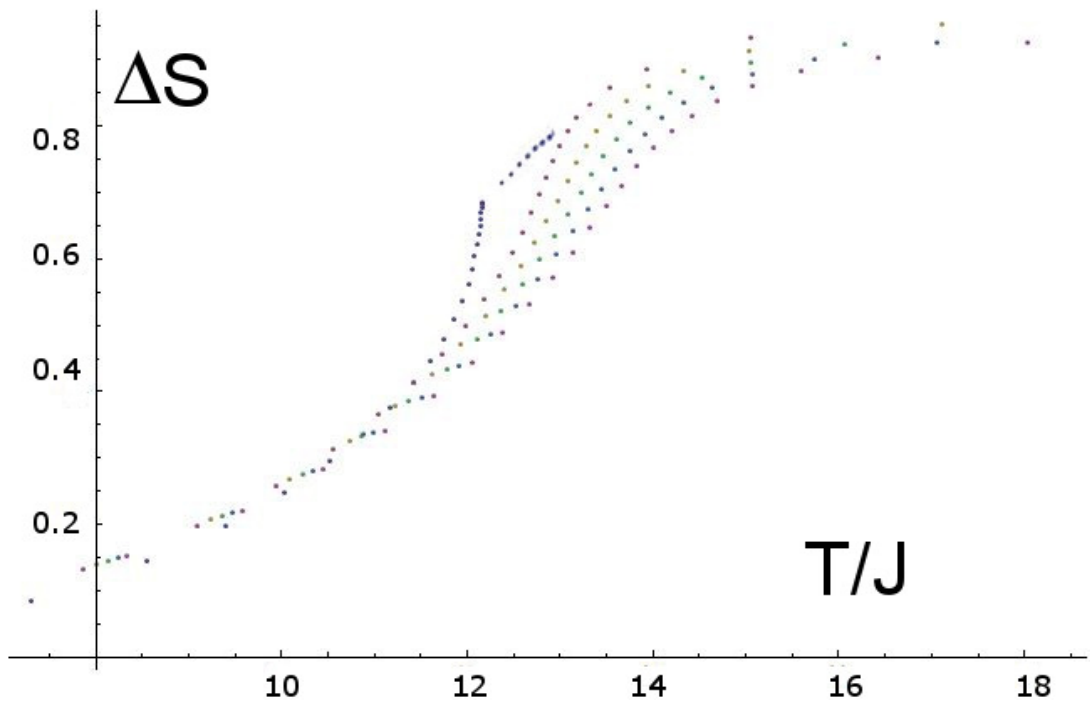

Figure 3

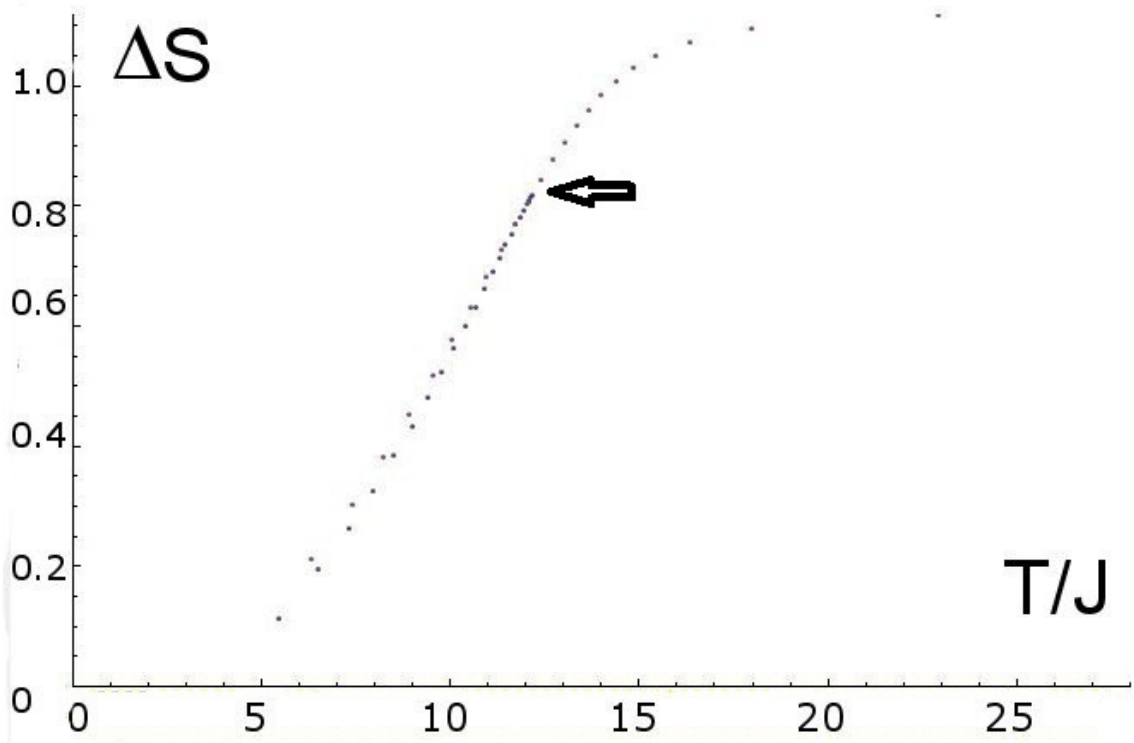

Figure 4

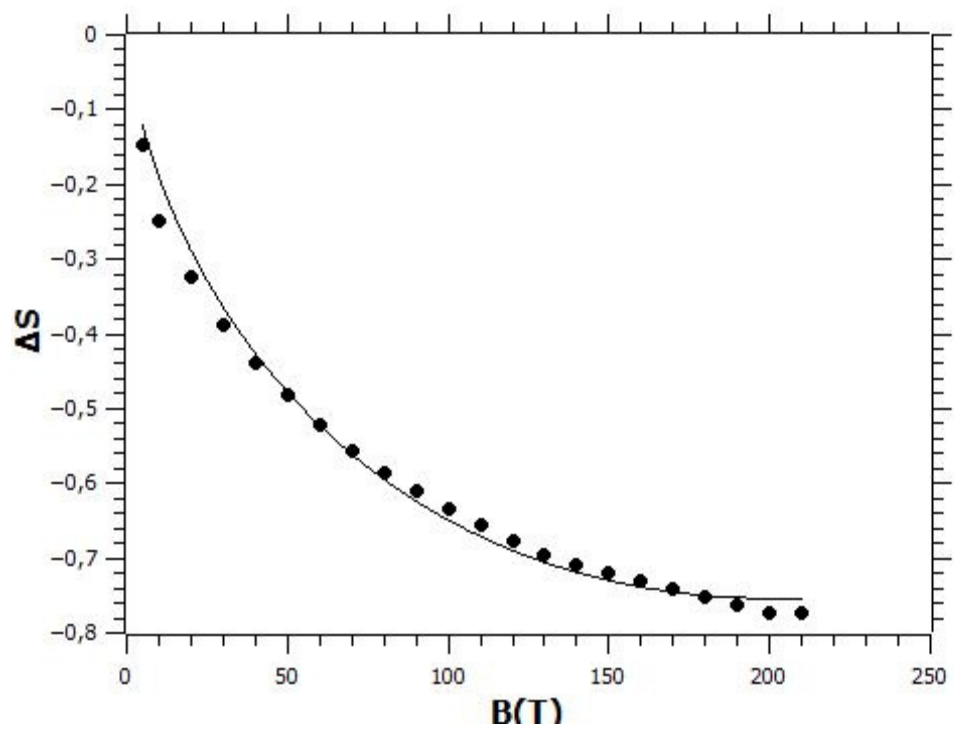

Figure 5

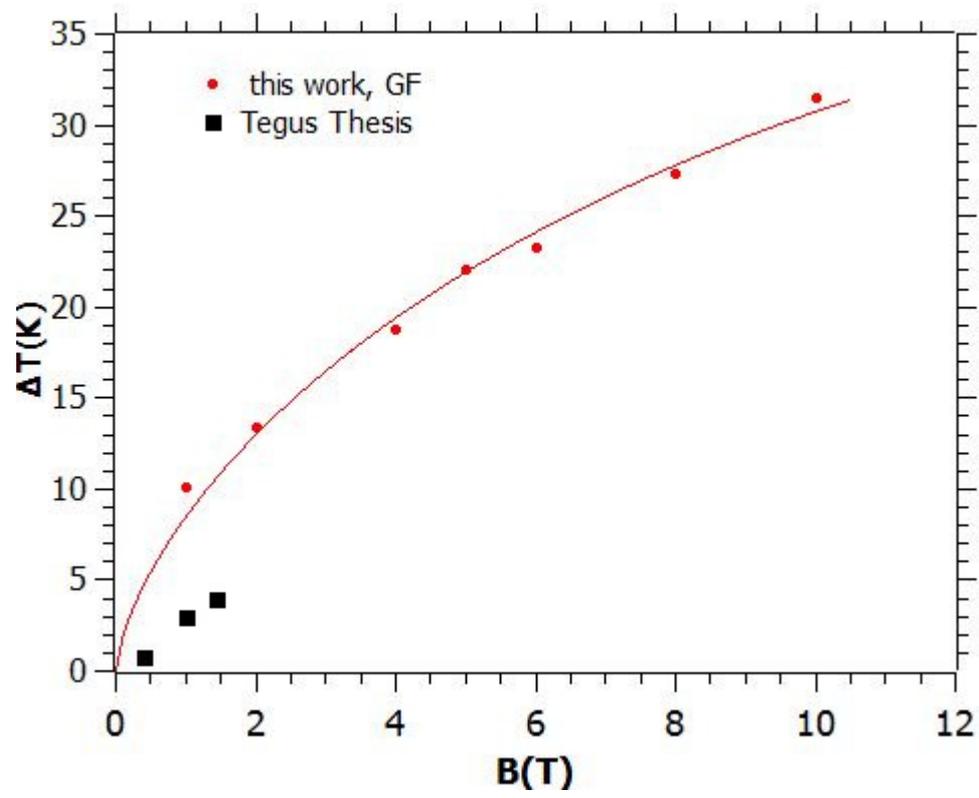

Figure 6

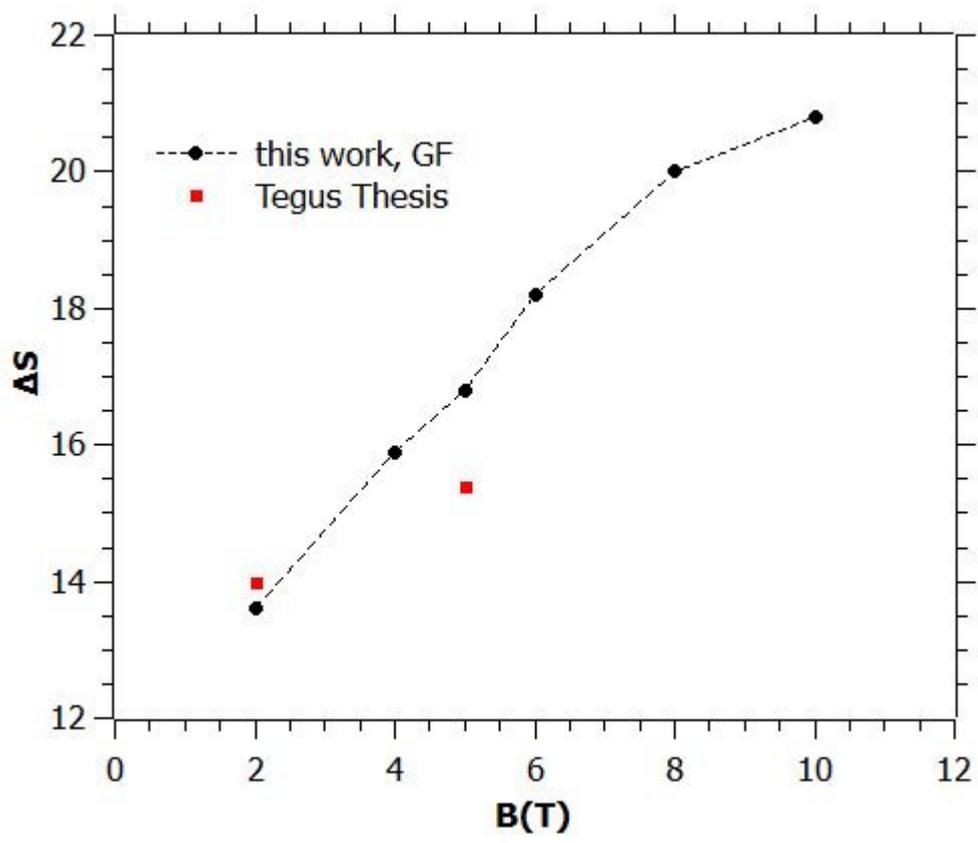